\documentclass[sigconf]{acmart}

\AtBeginDocument{%
  \providecommand\BibTeX{{%
    \normalfont B\kern-0.5em{\scshape i\kern-0.25em b}\kern-0.8em\TeX}}}
    
\copyrightyear{2021}
\acmYear{2021}
\setcopyright{iw3c2w3}
\acmConference[WWW '21]{Proceedings of the Web Conference 2021}{April 19--23, 2021}{Ljubljana, Slovenia}
\acmBooktitle{Proceedings of the Web Conference 2021 (WWW '21), April 19--23, 2021, Ljubljana, Slovenia}
\acmPrice{}
\acmDOI{10.1145/3442381.3450048}
\acmISBN{978-1-4503-8312-7/21/04}

\usepackage{soul}
\usepackage{cancel}

\hypersetup{
	colorlinks   = true, 
	urlcolor     = black, 
	urlbordercolor=black, 
	linkcolor    = blue, 
	linkbordercolor = white, 
	citecolor    = blue, 
	citebordercolor = white, 
	breaklinks   = true, 
}
\makeatletter
\Hy@AtBeginDocument{
	\def\@pdfborder{0 0 1} 
	\def\@pdfborderstyle{/S/U/W 0.5} 
}
\makeatother

\begin{document}
\raggedbottom

\addtolength\abovecaptionskip{-10pt}
\addtolength\textfloatsep{-15pt}

\title{Privacy Policies over Time: Curation and Analysis of a Million-Document Dataset}


\author{Ryan Amos}
\affiliation{%
  \institution{Princeton University}
  \country{United States}}

\author{Gunes Acar}
\affiliation{%
  \institution{imec-COSIC, KU Leuven}
  \country{Belgium}}

\author{Eli Lucherini}
\affiliation{%
  \institution{Princeton University}
  \country{United States}}

\author{Mihir Kshirsagar}
\affiliation{%
  \institution{Princeton University}
  \country{United States}}

\author{Arvind Narayanan}
\affiliation{%
  \institution{Princeton University}
  \country{United States}}

\author{Jonathan Mayer}
\affiliation{%
  \institution{Princeton University}
  \country{United States}}
%
\renewcommand{\shortauthors}{Amos, Acar, Lucherini, Kshirsagar, Narayanan, and Mayer}


\begin{abstract}
{Automated analysis of privacy policies has proved a fruitful research direction, with developments such as automated policy summarization, question answering systems, and compliance detection. Prior research has been limited to analysis of privacy policies from a single point in time or from short spans of time, as researchers did not have access to a large-scale, longitudinal, curated dataset. To address this gap, we developed a crawler that discovers, downloads, and extracts archived privacy policies from the Internet Archive’s Wayback Machine. Using the crawler and following a series of validation and quality control steps, we curated a dataset of 1,071,488 English language privacy policies, spanning over two decades and over 130,000 distinct websites. 

Our analyses of the data paint a troubling picture of the transparency and accessibility of privacy policies. By comparing the occurrence of tracking-related terminology in our dataset to prior web privacy measurements, we find that privacy policies have consistently failed to disclose the presence of common tracking technologies and third parties. We also find that over the last twenty years privacy policies have become even more difficult to read, doubling in length and increasing a full grade in the median reading level. Our data indicate that self-regulation for first-party websites has stagnated, while self-regulation for third parties has increased but is dominated by online advertising trade associations. Finally, we contribute to the literature on privacy regulation by demonstrating the historic impact of the GDPR on privacy policies.
}
\end{abstract}






\maketitle

\section{Introduction}
\label{sec:intro}
Privacy policies are one of the few available lenses for understanding how businesses interact with personal information. In the modern web ecosystem, many websites rely on monetizing user engagement as a primary revenue stream, which can reveal sensitive personal information to third parties such as advertisers and data brokers. Understanding privacy policies is, consequently, crucial for understanding both user privacy and economics on the web.

In this work, we aim to make privacy policies more accessible to researchers, who can in turn make privacy policies more useful to users, regulators, journalists, and other web stakeholders. There is extensive and valuable prior empirical work on privacy policies (Section \ref{sec:related}). Our goal is to advance privacy research by improving the scale and longitudinal scope of analysis.

\textbf{Dataset.} We built a crawler that discovers, downloads, and extracts text from privacy policies archived on the Internet Archive’s Wayback Machine. We used the crawler to assemble a privacy policy dataset that spans more than two decades and consists of over one million privacy policies from over 130,000 websites. The key challenge we faced was automatically identifying privacy policy pages and distinguishing them from similar documents at scale. We solved this problem in two steps. First, we developed a set of heuristics for downloading candidate privacy policies with high recall. The heuristics were based on manual analysis of hundreds of policies across several failure and success cases of our crawler. Next, to filter out non-privacy policies, we built a random forest classifier that achieves 98\% precision and 93\% recall. We describe a number of validation and quality control steps we performed throughout this process (Sections~\ref{sec:methods},~\ref{sec:classifier}, and~\ref{sec:dataset}).

Our dataset is publicly available and has received substantial attention, with over 80 access requests from private companies, industry research labs, and academic researchers as of the time of writing. 
To make our dataset more accessible, we have built a convenient change tracking interface (with GitHub as a backend) for examining how each privacy policy has evolved over time.

\textbf{Longitudinal analysis.} We used our dataset to conduct what is, to our knowledge, the longest-spanning and largest-scale longitudinal analysis of privacy policies (Sections~\ref{sec:doclevstatstime}~and~\ref{sec:analysis}).
Much of our analysis was guided by an automated trend detection tool that we developed to identify terms and concepts that may indicate shifts in the privacy policy landscape. Our system helped us identify trends related to self-regulatory organizations, third parties, tracking technologies, and regulations.
Our findings further undermine the idea that users can reasonably make informed decisions based on the information disclosed in privacy policies.

First, we find that over the last 20 years, privacy policies have become substantially longer---with a median length of 1,522 words in 2019---and moderately less readable, with the median policy being at a college reading level. Websites that are more popular have even less readable policies.

Second, we find that more than 20\% of websites have a privacy policy that links to one or more additional privacy policies. If a user wanted to understand the set of applicable privacy policies for one of these websites, they would face an even greater burden.

Third, our results show that, when compared to empirical measurements of web privacy practices, privacy policies vastly underreport tracking technologies and third parties.

Fourth, consistent with prior work on privacy regulation, we find that the GDPR prompted the most widespread changes to privacy policies in the last decade.

Finally, we examine which self-regulatory bodies have experienced growth and which have stagnated.

Our analysis and dataset help pave the way for researchers to understand the connection between government regulation, self-regulation, and user privacy. Understanding the impacts of past actions can help drive evidence-based methods for creating effective privacy policy legislation. Our work is a step in this direction.


{\section{Legal background}
\label{sec:background}

In this section, we review the evolving legal context for privacy policies in the United States and the European Union, since we aim to examine how privacy policies respond to legislative and regulatory developments. 

Since the early days of the commercial internet, privacy policies have had a hotly contested role in protecting user information. In the United States, voluntary disclosures about data practices by companies in their privacy policies are the foundation of the federal ``notice and choice'' approach to consumer privacy. The European Union has taken a different direction by specifying what should be included in privacy disclosures as part of its more comprehensive approach to data protection regulation. 

\textbf{Privacy policies in the United States: notice and choice.} In the mid-1990s, U.S. policymakers faced a fundamental question about how to regulate privacy online. They could directly regulate data practices, or they could leave privacy protections to the nascent online business ecosystem. Given evidence that market forces were failing to protect privacy and that this might be affecting economic growth, policymakers landed on a light-touch hybrid model. Online services would disclose their privacy practices to consumers on a mostly voluntary basis, then the Federal Trade Commission (FTC) would police those disclosures for accuracy~\cite{swire1997markets}. The FTC’s enforcement actions, which usually result in settlements with consent decrees, would provide guidance to the private sector about best practices and develop a common law of online privacy~\cite{solove2014ftc}.

The underlying theory behind the policy rested on ``the fundamental precepts of awareness and choice''~\cite{united1997framework}. Specifically, ``Data-gatherers should inform consumers what information they are collecting, and how they intend to use such data; and Data-gatherers should provide consumers with a meaningful way to limit use and re-use of personal information''~\cite{united1997framework}. Lurking in the background was the threat that policymakers would step in with regulation if the light-touch model failed. Additionally, Congress enacted legislation to require disclosures about data practices (and sometimes provide individual privacy rights) in specific sectors, including healthcare (the Health Insurance Portability and Accountability Act) and finance (the Gramm-Leach-Bliley Act).

\textbf{Children's Online Privacy Protection Act.} Children's privacy was the first internet-specific area to be directly regulated in the United States. In 1998, the FTC issued a study of children's privacy, which led it to recommend legislation that would place parents in control of their children's information. COPPA was enacted that year and regulates the collection and use of information of websites directed at children under 13 years of age. COPPA’s passage and the threat of further regulation led to the adoption of privacy policies on many popular websites.

\textbf{California Online Privacy Protection Act.} California introduced legislation that requires commercial websites to post privacy policies~\cite{caloppa}. While CalOPPA is a state law, after it came into force in 2004 it quickly became a de facto national privacy policy requirement. The recent California Consumer Privacy Act and California Privacy Rights Act~\cite{ccpaANDcpra} represent a break from the notice and choice model, guaranteeing consumers specific privacy rights.

\textbf{Privacy policies in the European Union: data protection.}
Europe has taken a more direct approach to privacy regulation, specifying the information that businesses must disclose about their privacy practices and guaranteeing individual privacy rights.

\textbf{Data Protection Directive.} The DPD, which went into effect in 1998, required EU member states to establish comprehensive privacy regimes that included disclosures, opt-out rights, and specific protections for sensitive data, including information about religious beliefs, sexual orientation, medical history, and financial circumstances. These rights were enforced by specialized regulatory agencies, data protection authorities, in the member states.

\textbf{General Data Protection Regulation.} In 2016, the EU overhauled its approach to protecting privacy by enacting the GDPR. The GDPR includes a comprehensive slate of privacy disclosure and control requirements, and it opens the door to serious penalties for violations. Most relevant to our work, the GDPR mandates a number of specific disclosures about how firms process individual data and how individuals can exercise their privacy rights. Online services typically provide these disclosures in their privacy policies.

}
\section{Related work}
\label{sec:related}

Increasing adoption of privacy policies and changes in the regulatory environment have led to a rich research literature. We briefly synthesize several strands of study that relate to this project.

\textbf{Marketplace and longitudinal studies.}
The earliest work on privacy policies consisted of marketplace surveys. A sequence of studies by the FTC~\cite{ftc-privacy-survey1998, ftc-privacy-survey2000} and Culnan~\cite{culnan2000protecting}, from 1998 to 2000, found that U.S. websites were rapidly adopting privacy policies but that the content of policies was often spotty. In the U.K., a 2002 Information Commissioner's Office (ICO) study reported similar results~\cite{ico-survey2002}. More recent surveys have called attention to privacy policy shortcomings in specific sectors, such as healthcare~\cite{sunyaev15} and finance~\cite{bowers2017}. In a market survey concurrent to this project, Srinath et al. report readability scores, topic models, key phrases, and textual similarity for a corpus of just over a million privacy policies~\cite{srinath2020}.

Several projects have expressly incorporated a longitudinal dimension. Milne and Culnan compared data from a set of marketplace surveys between 1998 and 2001, concluding that privacy policy adoption was increasing and that policies were including greater disclosures about collection, sharing, choice, security, and cookies~\cite{milne2002}. Later, Milne et al. compared a sample of privacy policies from 2001 and 2003, finding that readability was decreasing and length was increasing~\cite{milne2006longitudinal}. Antón et al. examined a small number of healthcare privacy policies between 2000 and 2003, observing similarly decreased readability and greater disclosures~\cite{anton2007hipaa}.

Recent work has used longitudinal privacy policy analysis to examine the effects of the GDPR. Degeling et al. used crawl data from 2018 and Wayback Machine data from 2016 and 2017 to examine privacy policies on over 6,000 websites before and after the GDPR took effect; the findings include a slight uptick in privacy policy adoption, increased use of key phrases related to the GDPR, and inconsistent privacy policy update practices~\cite{degeling2018we}. Linden et al. evaluated a privacy policy corpus of similar size, using Wayback Machine snapshots from 2016 and 2019, and found that privacy policies were longer, more likely to include categories of disclosures, and generally included more specific disclosures~\cite{linden2020privacy}.

We contribute to this area of literature with a longitudinal analysis of much larger scope, both in time (spanning over two decades) and in number of websites (over 130,000). We characterize longer-term document-level trends than the prior work, characterize trends by website popularity, and contextualize shifts associated with the GDPR (Section \ref{sec:doclevstatstime}). We also provide the first longitudinal analysis of specific trends (Section \ref{sec:analysis}).

\textbf{Research datasets.}
Another thread in the privacy policy literature is developing privacy policy research datasets to enable future study. Ramanath et al. contributed the earliest dataset in 2014, a collection of over 1,000 manually segmented privacy policies~\cite{ramanath2014unsupervised}. In 2016, Wilson et al. released OPP-115, a set of 115 manually annotated website privacy policies~\cite{wilson2016creation}. Companion work by Wilson et al. demonstrated the feasibility of crowdsourcing to annotate a privacy policy dataset~\cite{wilson2016crowdsourcing}. In 2019, Zimmeck et al. contributed APP-350, a dataset of 350 annotated mobile app privacy policies, and MAPS, a dataset of nearly 450,000 app privacy policy URLs~\cite{zimmeck2019}. Concurrent to this work, Srinath et al. released PrivaSeer, a dataset of over 1 million English privacy policies extracted from May 2019 Common Crawl data~\cite{srinath2020}. Also concurrent to this work, Zaeem et al. contributed a dataset of hundreds of thousands of privacy policies with associated website categories~\cite{zaeemlarge}.

Our work advances this area of the privacy policy literature by providing what is, to our knowledge, the first large-scale longitudinal dataset. We facilitate future work on changes over time to privacy policies by applying new techniques not only to current privacy policies but also to our curated dataset of historical policies.

\textbf{Compliance checking.}
A separate strand of research has examined whether privacy policies are in compliance with legal requirements and whether policies fully disclose data practices (regardless of legal requirements). Several projects have compared privacy policies to data flows and app permissions, noting pervasive gaps in disclosures~\cite{slavin2016, zimmeck2019}. Linden et al. compared pre- and post-GDPR privacy policies to ICO guidance and found improving though inconsistent compliance~\cite{linden2020privacy}. In the compliance work most similar to our own, Marotta-Wurgler examined nearly 250 privacy policies in 2016 for claims of compliance with regulatory and self-regulatory programs~\cite{marotta-wurgler2016}.

We contribute to the privacy policy compliance literature by evaluating representations about regulatory and self-regulatory programs over time. We identify the long-term rise of specific self-regulatory programs,
 and we identify shifting representations about regulatory programs in response to legal developments
(Section \ref{subsec:trends}). We also compare aggregate privacy policy disclosures about advanced tracking technologies to the known level of adoption of those technologies (Section \ref{subsec:trends}).

\textbf{Consumer comprehension.}
Another line of privacy policy research examines consumer comprehension. Surveys and intervention studies have demonstrated that consumers inconsistently read privacy policies and primarily read privacy policies to exercise control~\cite{milne2004}; consumer trust in websites has little connection to their privacy policies~\cite{pan2006}; consumers have limited comprehension of privacy policy content~\cite{vail2008, mcdonald2009comparative}; consumers interpret privacy policies differently from legal experts~\cite{reidenberg2016disagreeable, strahilevitz2016}; and consumers have difficulty exercising privacy choices in privacy policies~\cite{habib2020}. Studies have repeatedly shown that privacy policies are lengthy and difficult to read~\cite{jensen2004, mcdonald2008cost, li2012online, fabian2017large}. While the precise readability metrics vary by project, recent work has demonstrated that these metrics are strongly correlated for privacy policies~\cite{fabian2017large}. Textual analysis has also highlighted the prevalence of ambiguous claims in privacy policies~\cite{reidenberg2016ambiguity}.

We advance the literature on consumer comprehension of privacy policies by tracking how length and readability have evolved over two decades and across website ranking tiers.

\textbf{Applied machine learning.}
A final related area of the privacy policy literature applies machine learning techniques to 
privacy policy texts. Recent work has identified specific claims in policies and assigned overall grades~\cite{zimmeck2014}, identified choices provided in privacy policies~\cite{sathyendra2017, kumar2020}, and enabled structured and free-form queries about privacy policy attributes~\cite{harkous2018polisis}. While our own privacy policy analysis relies on heuristics rather than  machine learning, the dataset we contribute is intended to enable future machine learning projects to easily present longitudinal results.
\section{Collecting historical privacy policies}
\label{sec:methods}
Privacy policies are intended for human readers, and automating the process of locating the policies and extracting their text is challenging. A further complication is the historical dimension: our goal is to extract privacy policies of websites that may not exist today.
In this section, we explain how we addressed these challenges. We discuss how we selected our target set, ethically crawled policies, and ensured a high level of quality.

\begin{figure*}[t]
\centering
\includegraphics[width=0.99\textwidth]{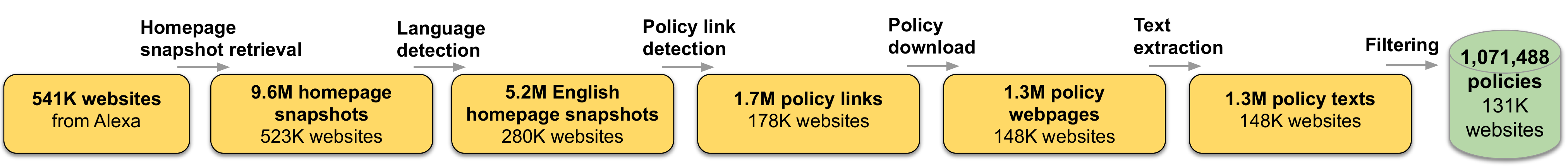}
\caption{Overview of the data collection steps.}
\Description[Data Collection Pipeline]{A flow diagram for our data collection pipeline. First we collect 541K websites from Alexa. Then we collect 9.6M homepage snapshots from 523K of those websites. We reduce this to 5.2M English homepage snapshots from 280K websites. We find 1.7M documents with privacy policy links from 178K websites. We find 1.3M privacy policy webpages from 148K websites. We extract text from 1.3M of those. Then we filter, leaving us with 1,071,488 policies from 131K websites.}
\label{fig:data-collection}
\end{figure*}

\subsection{Building the list of target websites}
\label{sec:sec:building-website-list}
We began by identifying a set of websites to include in our data collection process.
We chose websites that appeared in the Alexa top 100K list between 2009 and 2019. We considered 100K websites to be sufficient to reach into the long tail of the web, yet not so numerous as to pose computational challenges for data collection or analysis.

Next, we selected a method for discretizing time windows in our longitudinal dataset. We struck a trade-off between enabling granular analysis and limiting computational and storage requirements. For each website, we retrieved one snapshot from the first half of the year (January-June), and one from the second half (July-December). We call these six-month time spans~\emph{intervals}, and we use intervals as the basic time unit of our data collection and analysis. We refer to the first interval in a year as ``A'' and the second as ``B,'' so the first half of 2005 is ``2005A.''
For each interval, we used the daily archives of the Alexa top million list~\cite{naab2019prefix} to retrieve the Alexa rankings closest to the interval's midpoint: either March 31 (A) or September 30 (B).
We collected the Alexa ranks for each interval from the beginning of Alexa’s publication in 2009 to 2019.
We obtained 541,616 websites by combining all domains that appear in the top 100K of these 22 Alexa lists (two lists per year for 11 years).

\subsection{Building the list of snapshots}
\label{sec:sec:snapshot-retrieval}
Next, we determined the list of homepage snapshots available on the Wayback Machine
by querying their \emph{CDX Server API}~\cite{wayback-cdx-api}.
When a website had multiple snapshots in a six-month interval, we picked the snapshot closest to the midpoint of the interval.
We did not restrict our queries to a time window when searching for snapshots.
For instance, if {\tt example.com} was only listed in the Alexa top 100K in 2019,
we included its snapshots from any other year including as early as 1996, the first year for which Internet Archive has crawls~\cite{WaybackMachineGeneralInformation}.

\subsection{Downloading privacy policies}
\label{sec:sec:download-policies}
To download archived privacy policies, we built a custom crawler based on 
\emph{Pyppeteer}~\cite{miyakogi2019May}.

{\textbf{Language detection.}}
We limited ourselves to privacy policies in English, because our work is motivated by U.S. and EU developments and because evaluating privacy policies in other languages would require additional language proficiency. To exclude non-English websites from our crawls, we ran a language detection crawl that loaded the most recent homepage snapshot of each website, extracted the page text and identified the language of the extracted text using the \emph{Polyglot} Python library~\cite{polyglot-pypi}. If the latest snapshot failed to load, we tried to visit up to three random homepage snapshots to identify the site's language. 
We identified 280,798 English websites (5,223,228 snapshots),
and discarded the remaining 243,060 websites.

{\textbf{Loading the homepage snapshot.}}
To download archived privacy policies, we first loaded the homepage snapshot and ran a language check to make sure the snapshot was in English to account for changes in ownership or localization.
Second, we aborted visits where the crawler attempted to load the live (non-archived) version of the page or a snapshot from another interval due to a redirection.
Finally, we monitored all requests that the crawler made and blocked requests that attempted to fetch resources from live websites.

{\textbf{Privacy policy link detection.}}
Privacy policy links are not universally standardized in their text or URL.
We chose to use \textit{link texts}---that is, the clickable text appearing within the \texttt{<a>} element---to detect privacy policy links over other link features such as the URL path. Link texts are expected to be recognizable by users, and therefore more likely to contain certain keywords.

We detected policy links using exact and partial keyword matching.
We compiled these terms based on prior research~\cite{libert2018automated} and adding other terms by manually analyzing a sample of 100 pages where we did not find a privacy policy link in a pilot crawl. 
The manual analysis of these pages involved searching for privacy policy links and checking whether the linked page’s title, headers and content described a privacy policy or not. 

The link detection method is designed to be comprehensive and may lead to pages that do not contain actual privacy policies.
We detected and removed these false positives after the crawl (Section~\ref{sec:classifier}).

{\textbf{Policy download.}}
For each detected privacy policy link, we queried the Wayback Machine CDX API to retrieve the list of snapshots that were in the same time interval as the homepage snapshot. 
If the link URL ended in ``.pdf'', we used the Python \textit{requests} library~\cite{PyPI-requests} to retrieve the document.
If the link URL did not end in ``.pdf,'' we loaded the policy snapshot URL using the Pyppeteer-based crawler and ran a final language check to eliminate non-English policies.

{\textbf{Boilerplate removal and text extraction.}}
Privacy policy webpages typically include \emph{boilerplate} content that is separate from the policy (e.g., sidebars, footers, and headers). 
We removed boilerplate and extracted the main article from the page's DOM using the standalone version of Mozilla’s \emph{Readability} library~\cite{readability-mozilla}. 
Next, we extracted Markdown formatted text from the \emph{readable} policy web page using the \emph{html2text} Python library~\cite{html2text-github}. Markdown formatting allowed us to retain the links and basic document structure such as headers and lists, with minimal markup overhead.

\subsection{Practical and ethical considerations}
\label{subsec:practical-matters}
We confirmed that the Internet Archive’s terms of use do not prohibit automated access. Prior to starting our crawls, we sent an email to the IA’s contact address to notify them of our study.
We note that several prior works used Wayback Machine~\cite{lerner2016internet, lerner2017rewriting,brunelle2015not,brunelle2016impact}.

We took steps to minimize any adverse impact of our crawl on the Wayback Machine’s servers.
To reduce our bandwidth footprint, we disabled image downloads and limited the number of parallel crawl workers to 256.
We slowed down our crawls by pausing the workers when they received HTTP 429 (``Too Many Requests'') or HTTP 503 (``Service Unavailable'') errors from the Wayback Machine. 
After starting the crawl, we monitored the Wayback Machine server load stats~\cite{Wayback-Stats} to ensure that we were not imposing a significant load.

\subsection{Evaluating data quality}
\label{subsec:failure-analysis}

To ensure our dataset was of high quality, we manually investigated causes of failure (i.e., homepages where the crawler did not extract privacy policy text). We started with 5,223,228 homepage snapshots, and our crawler was able to download 1,292,420 privacy policy snapshots (24\%).
Downloads of privacy policies corresponding to the other 3,930,808 homepage snapshots failed due to various causes which we list in Table~\ref{tab:failure-cause}. While the high frequency of absent policies may be counter-intuitive, we found that only about 2\% of missing policies were attributable to crawler limitations.

The most common cause for a missing privacy policy, by far, was the crawler loading an archived homepage but failing to identify a privacy policy link.
The next most common cause was a~\emph{blank homepage}, where the crawler could not extract any text. Another recurring issue was that, though we had attempted to filter our non-English sites before the crawl, some homepage snapshots were classified as non-English.

To investigate the root causes of crawler failures, we manually analyzed 100 random snapshots for each of the most common crawl failures.
The main takeaways are:
only 4 of 100 homepages where the crawler did not identify a privacy policy link actually had a privacy policy link. Only 3 of 100 homepages detected as \emph{blank} contained a privacy policy link, while 13 contained some text in a separate frame--- typically old webpages. We further explore the relation between the snapshot age and crawl success in 
Section~\ref{subsec:data-overview}.

In sum, the overwhelming majority of crawl failures were attributable to homepage snapshots that did not contain a policy link, were not archived by the Wayback Machine (e.g., due to \texttt{robots.txt} rules), or were in a language other than English. The missing privacy policies that are attributable to limitations of the crawler are about 2\% of all snapshots (4\% of 44.7\% + 3\% of 7.3\%).

\begin{table}[]
\centering
\resizebox{0.9\columnwidth}{!}{%
\begin{tabular}{@{}lrr@{}}
\toprule
\textbf{Failure cause}                           & \textbf{Count} & \textbf{Percent} \\ \midrule
No privacy policy link found on homepage         & 2,336,849      & 44.7\%       \\
Homepage detected as blank                        & 383,337       & 7.3\%      \\
Non-English homepage                             & 310,563        & 5.9\% \\
Policy page is not archived in this interval     & 271,736        & 5.2\%\\
Out-of-interval redirection for homepage         & 158,727        & 3.0\%\\
\bottomrule
\end{tabular}%
}
\caption{The five most common causes for the crawler to not download a privacy policy for a homepage. Out-of-interval redirection (row 5) means our crawler was redirected by the Wayback Machine to a snapshot in a different interval.}
\label{tab:failure-cause}
\end{table}

\section{Filtering the corpus}
\label{sec:classifier}

As discussed in Section \ref{sec:sec:download-policies}, we expect that some of the downloaded pages are not privacy policies. In order to filter out web pages that are not privacy policies, we trained a classifier on manually labeled data from the candidate policies. In this section, we refer to the text of the crawled web pages extracted with \emph{html2text} as \textit{documents}.

\textbf{Labeling definition.}
\label{sec:sec:pp-definition}
In order to label documents as privacy policies (or not), we needed to adopt a definition of what constitutes a privacy policy. We did not select a definition based on legal requirements, since both privacy policy text and applicable law can be ambiguous; we considered it outside the scope of the project to develop a classifier that accurately determines whether a document satisfies relevant legal requirements. Moreover, different jurisdictions have different legal requirements, which would necessitate further judgment about the applicable law for a given website.

Instead, we aimed to label whether a document is an \textit{apparent} privacy policy, without determining whether the document meets any particular legal requirements. Specifically, we labeled a document as a privacy policy if it met all of the following criteria:
\begin{itemize}
    \item The document relates to privacy. This criterion eliminates documents that are irrelevant, notwithstanding the link text.
    \item The document is legal in nature (e.g., uses legal terminology or appears to have legal significance). This eliminates documents that informally discuss privacy, such as blog posts.
\end{itemize}
In the course of applying our criteria, we encountered several recurring borderline types of documents. We labeled the following categories of documents as not privacy policies, because the contents were dissimilar or could otherwise be problematic for automated textual analysis.
\begin{itemize}
\item Incomplete privacy policies.

\item Cookie policies, which only describe the privacy practices of the website with respect to cookies.

\item Security policies, which only describe the security practices of the website.

\item Mixed documents containing paragraphs that are unrelated to privacy, such as a privacy policy intertwined with terms of service.

\item Privacy centers or landing pages, which further link to a privacy policy.

\item Privacy policy summaries, which further link to a long-form privacy policy.

\end{itemize}

\textbf{Labeling the sample.} We started by manually labeling a sample of candidate privacy policies (i.e., documents returned by the crawler) using our criteria. First, we randomly sampled and labeled 1,139 documents. In the sample, the majority of the documents 
matched the ``privacy + policy'' link text pattern (as described in Section \ref{sec:sec:download-policies}). We additionally sampled 50 random documents for each of the six remaining link text patterns so as to reduce overrepresentation of the 
dominant
link text pattern.

To confirm that the labeling was consistent with our criteria, two researchers independently labeled a sample of 100 randomly collected documents using the criteria above. The inter-annotator agreement was $\kappa=0.93$.

Our final labeled dataset contains 1,173 positives (i.e., documents that are privacy policies) and 266 negatives (i.e., documents that are not privacy policies), respectively $\sim$81.5\% and $\sim$18.5\%.

\textbf{Converting documents to features.} We cleaned the documents by removing Markdown tags and retaining inner text. We then removed symbols and the set of English stop words in \emph{scikit-learn}~\cite{scikit-learn}. Next, we created word n-gram tokens spanning from unigrams to 4-grams and generated features based on the frequency of occurrence of each token. We excluded tokens that appeared in fewer than 1\% of the documents.

We also extracted features from document titles, parsing them from the HTML \texttt{<title>} tag. For PDF files, we obtained titles by using the \textit{pdftitle} library~\cite{pdftitle-pypi}.
We removed symbols and stop words from the titles and we generated features based on the number of occurrences of each unigram in the title. We excluded tokens that appeared in fewer than 1\% of the documents.

We additionally extracted features from the link text and document URL. We found that these features did not contribute to classification performance, so we excluded them from our model.

\begin{figure}[]
\centering
\includegraphics[width=0.8\columnwidth]{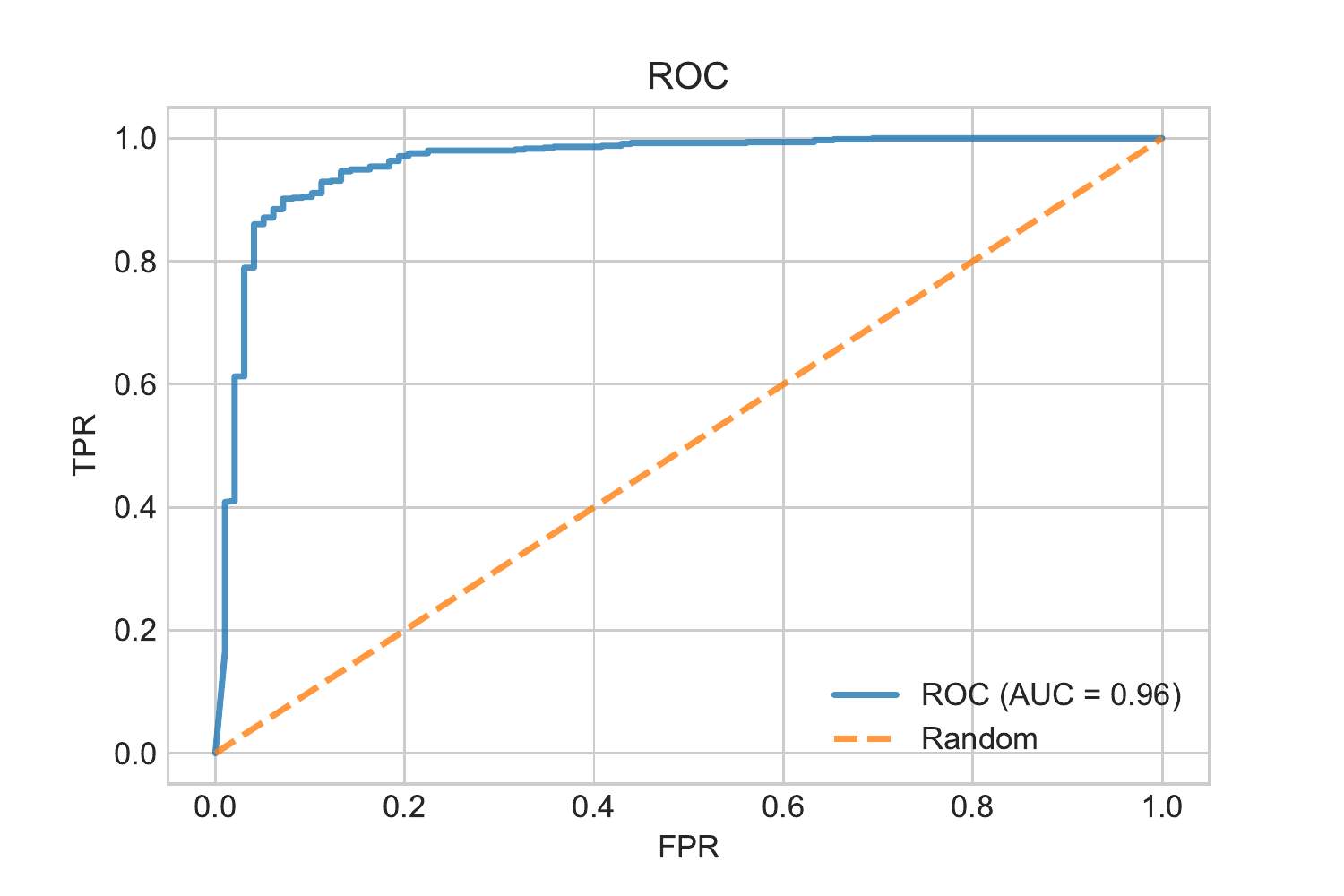}
\caption{Predictive performance of the privacy policy random forest classifier, applied to held-out documents.}
\Description[ROC curve]{An ROC curve, with an AUC of 0.96}
\label{fig:roc-auc}
\end{figure}

\textbf{Model training and validation.}
We trained and tested two model types, random forest and logistic regression, using 10-fold cross validation across varying hyperparameters. We evaluated the models based on maximizing average area under the receiver operating characteristic curve (\textit{AUC}), since that metric accounts for the class imbalance. The best performing random forest model had a higher mean AUC than the best performing logistic regression model (97\% vs. 95\%), so we selected the random forest model.

We then considered the tradeoff between precision and recall for the random forest model. We manually selected a classification threshold that prioritized precision over recall, since we anticipate uses of our corpus (e.g., many forms of automated analysis) that may be more sensitive to erroneously including non-privacy policy documents than erroneously omitting privacy policies.

Finally, we evaluated our model on a held-out set of 749 randomly sampled and manually labeled documents, consisting of 651 positives and 98 negatives. The performance of the final model on the held-out set is shown in Figure \ref{fig:roc-auc}. The chosen classification threshold resulted in 97.9\% precision and 94.2\% recall.

\textbf{Document Classification.}
Approximately 17\% (220,932) of the documents in the dataset were classified as negatives, with the final dataset totaling 1,071,488 policies. We discuss the release of this data in Section \ref{sec:conclusion}.

\newcommand{\absentPerc}{80.0} %
\newcommand{\absentEngPerc}{61.3} %
\newcommand{\absentEngOneKPerc}{48.8} %
\newcommand{\gapsPerc}{81.8} %
\newcommand{\pregapsPerc}{76.5} %
\newcommand{\midgapsPerc}{7.4} %
\newcommand{\postgapsPerc}{16.1} %

\newcommand{\midgapsPossPerc}{29.8} %
\newcommand{\midgapsDomainsPerc}{54.1} %
\newcommand{\lifespanNoGaps}{11.2} %
\newcommand{\lifespanGaps}{16.5}

\newcommand{\numpresentdoms}{108,499} %
\newcommand{\numabsentenglishdoms}{150,192}

\newcommand{\midgapNoRankPerc}{3.0}
\newcommand{\pregapNoRankPerc}{78.4}
\newcommand{\postgapNoRankPerc}{11.2}
\newcommand{\allgapNoRankPerc}{92.7}
\newcommand{\notgapNoRankPerc}{7.3}

\newcommand{\percSnapshotWithPrior}{85.3} %
\newcommand{\percSnapshotWithoutPrior}{14.7} %

\newcommand{\corrMidgap}{-0.02} %
\newcommand{\corrPregap}{0.07} %
\newcommand{\corrPostgap}{0.08} %

\newcommand{\nSnapshots}{910,546}
\newcommand{\nSites}{108,499}

\section{Corpus composition}
\label{sec:dataset}

In order to understand the structure and biases of the data, we discuss the dataset composition.

\subsection{Cleaning the corpus}
\label{sec:deduplication}
In order to ensure the quality of our analysis, we purged additional, real policies from our dataset that might lead to a biased analysis. In total, we removed 160,941 policies following the steps listed below, leaving us with 910,546 policies from 108,499 sites. We call the resulting corpus the \textit{analysis subcorpus}. In sections \ref{sec:dataset}, \ref{sec:doclevstatstime}, and \ref{sec:analysis} we use the analysis subcorpus for our analyses.\footnote{While we removed this data for our analyses, we anticipate that there may be some research questions that would benefit from their inclusion. Accordingly, in our public release, we include both the full corpus and the analysis subcorpus.} 

\textbf{Parked domains.} As our dataset spans more than 20 years,
many domains have expired and were parked for monetization.
We found that privacy policies of the parking services such as {\tt godaddy.com} and {\tt hugedomains.com} were abundant in our dataset as they hosted privacy policies of 7,092 and 2,275 distinct domains, respectively.
We built a detection method based on the lists of domain parking services and registrars from prior research~\cite{vissers2015parking, kuhrer2014paint, kuhrer2014paint-TR, wang2006strider},
and the ICANN-Accredited Registrars list~\cite{ICANN-Registrars}.
With this method, we detected 79 parking providers and resellers and removed 24,289 (2.3\%) snapshots hosted by these providers.
 
\textbf{Homepage redirection.} In order to ensure the metadata is consistent with policy content, we removed policy snapshots with a cross-origin homepage redirection (COHR), unless both sites were likely owned by the same entity. Thus, we removed COHR policy snapshots, but we made an exception for websites with
common~\emph{label}s in their domain names (e.g. \texttt{google.com} and \texttt{google.ca})
as this may indicate that they are owned by the same organization.

\textbf{Policy URL.} Finally, to remove duplicates, we followed
Degeling et al.'s approach based on privacy policy URLs~\cite{degeling2018we}. When two or more sites shared a policy URL during the same interval, we kept only one, preferring the snapshot where the homepage domain matched the privacy policy domain. This removed 61,220 policies.

\subsection{Corpus overview}
\label{subsec:data-overview}

\begin{figure}[]
\centering
\includegraphics[width=0.99\columnwidth]{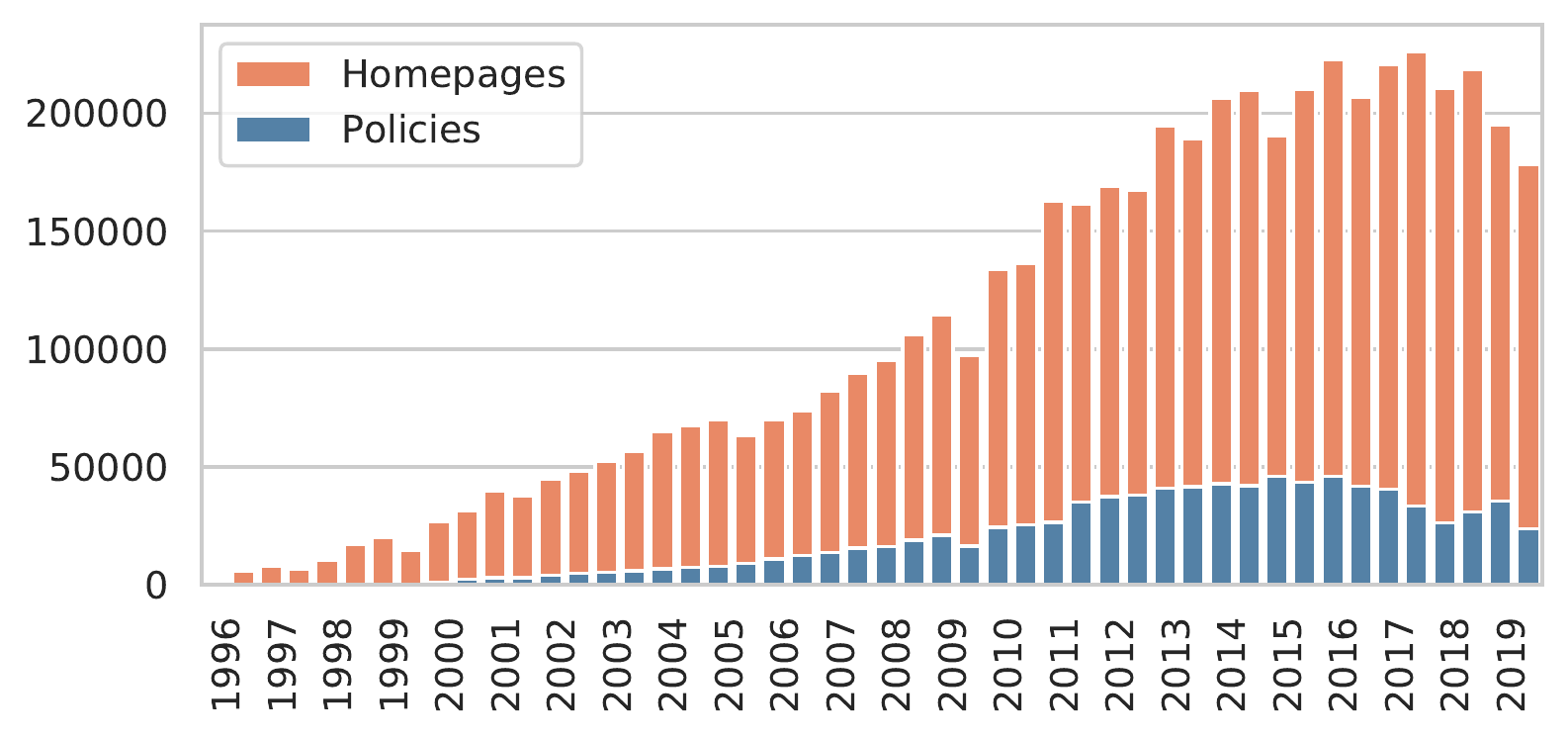}
\caption{The number of homepage snapshots and privacy policies for each interval. Note that each bar represents one interval and there are two intervals per year.}
\Description[Bar plot of homepages versus policies, by year]{Generally, there are many more homepages than policies. The number of policies peaks in 2015 but declines slightly. The number of homepages roughly matches the same trend as the number of policies. }
\label{fig:numpolicies}
\end{figure}

On average, a website has 8.4 privacy policy snapshots ($M=6$). Although the dataset contains policies from as early as 1997, 79.4\% of the policies are from snapshots archived in 2010 or later.
As we briefly explored in Section~\ref{subsec:failure-analysis}, we expected the rate of successful downloads from earlier snapshots to be lower, in part due to different website structures, such as greater use of frames.

Figure~\ref{fig:numpolicies} shows the number of privacy policies and homepage snapshots per interval.
The decreases in the number of policies in 2009B and 2018A also occur in the number of homepage snapshots in those intervals, indicating that these decreases are due to changes in archiving or incomplete archives for those intervals.

\begin{figure}[t]
\centering
\includegraphics[width=1\columnwidth]{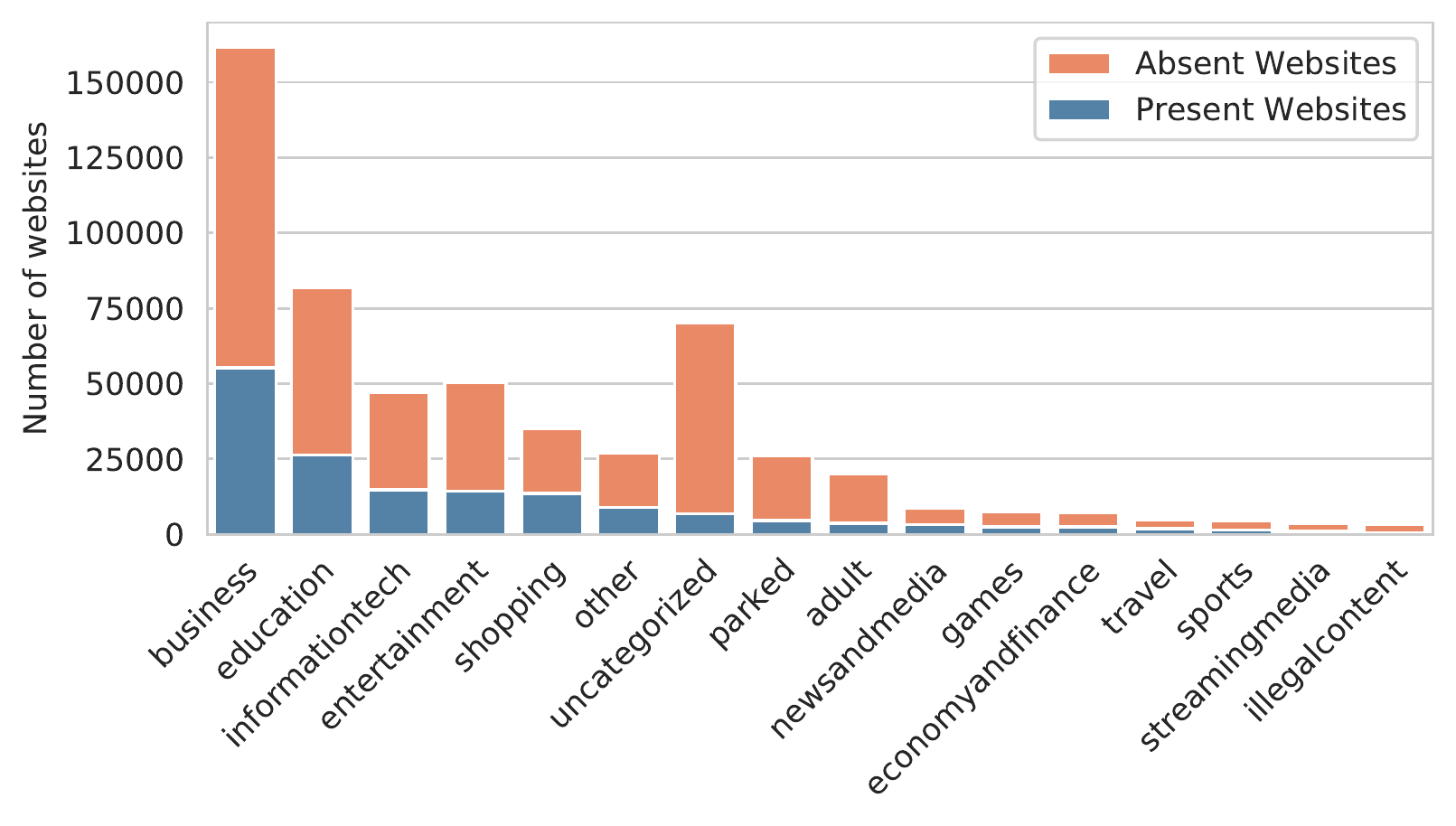}

\caption{The distribution of website categories for present and missing sites. ``Other'' is composed of the 27 least frequent categories for English language websites. Websites which belong to multiple categories are counted once per category. Websites with no listed categories are added to the ``uncategorized'' category.}
\Description[Histogram of category vs number of websites]{This diagram shows the number of websites that we tried to capture and that we succeeded in extracting an English language policy from. Categories in descending order: business, education, informationtech, entertainment, shopping, other, uncategorized, parked, adult, newsandmedia, games, economyandfinance, travel, sports, streamingmedia, illegalcontent.}
\label{fig:cat_dist}
\end{figure}

{\textbf{Distribution of website categories.}}
We examined the website categories for which we have the best coverage. We collected category data from Webshrinker~\cite{Webshrinker}, which provides a domain category lookup API.
Because category data is not available historically from Webshrinker, we assumed that categories are constant across time. We collected category data for the \numpresentdoms~websites in our dataset (present) and an additional \numabsentenglishdoms~websites with English homepages that were at one point in the Alexa top 100K (absent), which we show in Figure~\ref{fig:cat_dist}.
Just a few categories dominate both our dataset and English websites in general, and uncategorized websites are strongly underrepresented in our dataset.

{\textbf{Distribution of snapshot Alexa ranks.}}
The distribution of homepages and snapshots by Alexa rank is shown in Table~\ref{tab:rate-of-download-per-rank-bin}.
The >1M bin contains snapshots for which we do not have an Alexa rank, either because they were not listed in the Alexa top 1M or were captured
before 2009.
While the majority of our policies come from websites with rank 100K and up, the success rate of obtaining a policy is much higher if the website is in the top 10K. 

\begin{table}[]
\centering
\resizebox{0.9\columnwidth}{!}{%
\begin{tabular}{@{}lrrr@{}}
\toprule
\textbf{Alexa ranks} & \textbf{Homepage snapshots} & \textbf{Privacy policies} & \textbf{\%} \\ \midrule
(1, 1K{]}       & 13,455   & 5,003   & 37.2 \\
(1K, 10K{]}     & 104,801  & 38,959  & 37.2 \\
(10K, 100K{]}   & 980,928  & 278,324 & 28.4 \\
(100K, 1M{]}    & 1,339,157 & 319,866 & 23.9 \\
\textgreater 1M & 2,786,853 & 268,394 & 9.6  \\ \bottomrule
\end{tabular}%
}
\caption{Rate of successful privacy downloads per Alexa rank buckets -- based on privacy policies in the analysis subcorpus.}
\label{tab:rate-of-download-per-rank-bin}
\end{table}

\newcommand{\medianWCStartAlexa}{876}
\newcommand{\medianWCEnd}{1522}
\newcommand{\medianFKStart}{11.9}
\newcommand{\medianFKEnd}{13.2}

\section{Document-level trends}
\label{sec:doclevstatstime}

We examine how length, readability, and updates have changed over time. Because 
most of the privacy policies in our corpus are
from 2009-2019, we focus most of our analysis on that time range.

\subsection{Comprehension}
\begin{figure}[t]
\centering
\includegraphics[width=0.99\columnwidth]{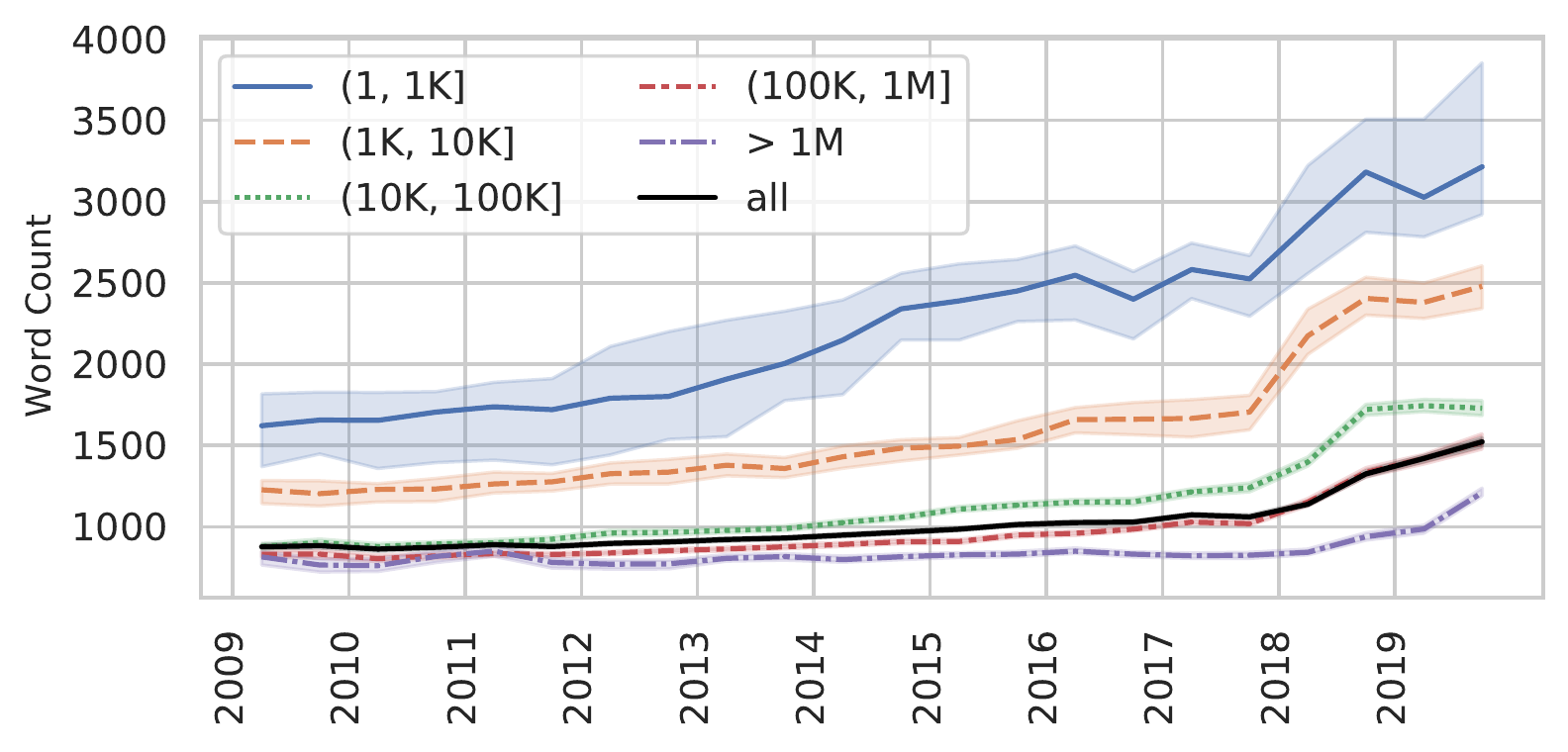}
\caption{The median word count of policies binned by Alexa rank. The highlighted region in this and the following figures shows the 95\% confidence interval. Ranks are at the time of the snapshot.}
\Description[Line plot of year vs word count]{The trend is generally slightly upwards and a similar shape across bins. Bins closer to 1 have more words.}
\label{fig:wordcount}
\end{figure}

{\bf Policy length.} Figure~\ref{fig:wordcount} shows the change in policy length as measured by word count. The median word count has increased gradually over time --- doubling between 2009A (\medianWCStartAlexa) and 2019B (\medianWCEnd) --- and more sharply in recent years, after the introduction of the GDPR. This trend holds for both popular and less popular websites. But the median hides a substantial variance: the 5\textsuperscript{th} percentile for word length is 248 words and the 95\textsuperscript{th} percentile is 3404 words. 
Our measurements for the top 10K are roughly consistent with the March 2016 measurements study by Degaling et al., but less so for their post-GDPR March 2018 measurements. Degaling et al. found the median privacy policy word count of the top 500 websites for 28 EU countries to be 2,145 in 2016 (our data: 1K: 2,691, 10K: 2,122) and 3,044 in 2018 (our data: 1K: 3,303, 10K: 2651) ~\cite{degeling2018we}. We believe the inconsistency is due to the difference in website sampling methods.

{\bf Readability.}
\label{sec:readability}
Several studies have shown that privacy policies are difficult to read~\cite{mcdonald2008cost,fabian2017large,milne2006longitudinal,li2012online}. 
Here we examine how readability has changed over time.
\begin{figure}
\centering
\includegraphics[width=1\columnwidth]{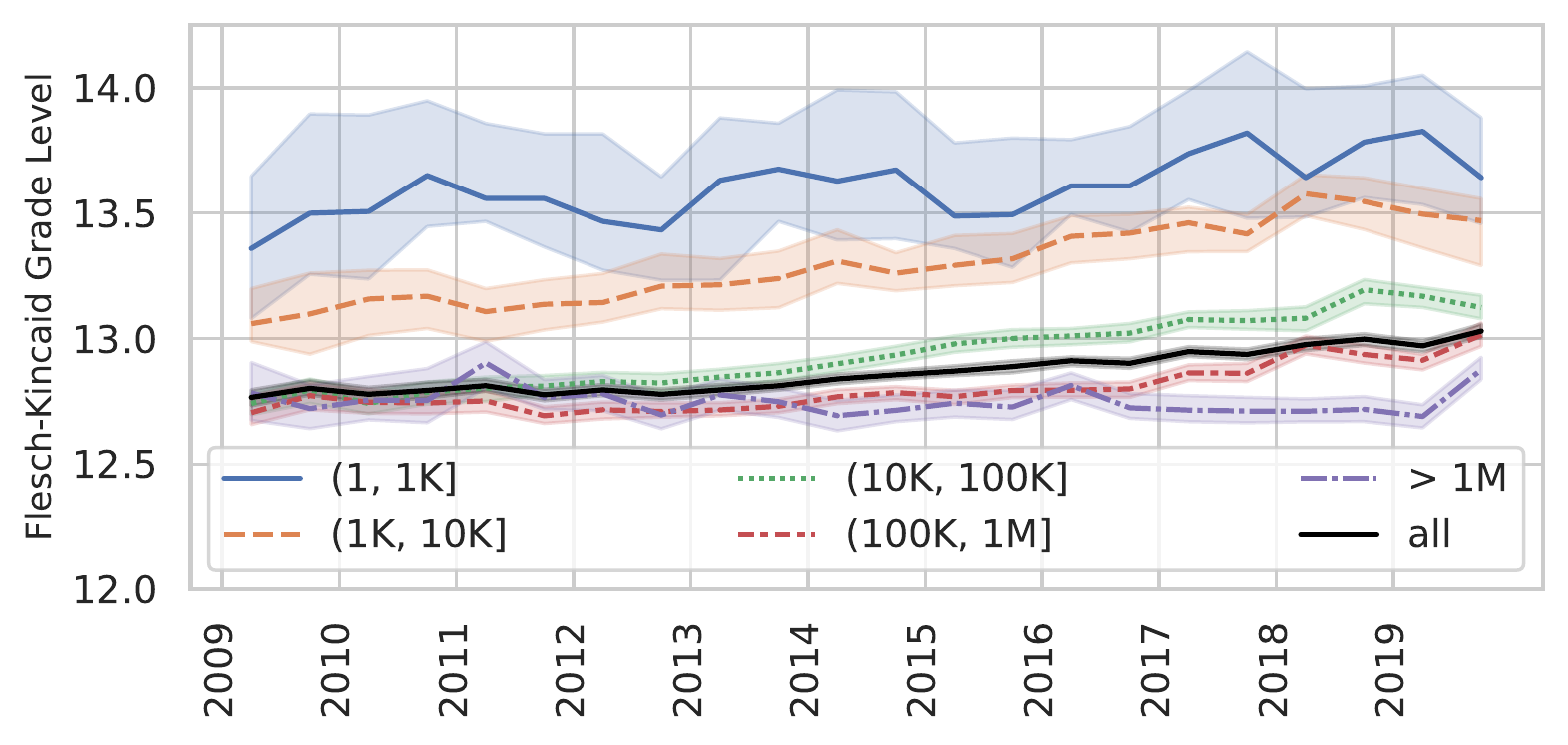}
\caption{Median Flesch-Kincaid grade level from 2009 to 2019, binned by Alexa rank. 
}
\Description[Line plot of year vs Flesch Kincaid Grade level]{The trend is generally flat. Bins closer to 1 have a higher reading level.}
\label{fig:readability}
\end{figure}
We measured readability with the Flesch-Kincaid grade level (FKGL)~\cite{kincaid1975derivation}
using the \textit{py-readability-metrics} Python library~\cite{pyreadabilitymetrics}.
Our preprocessing involved removing non-sentence text such as headers, tables, and lists by writing a custom Markdown renderer using the \emph{Mistune}~\cite{mistune} Python library.

The FKGL measurements are shown in Figure~\ref{fig:readability}. The median readability score has risen more than a grade level from 2000A (\medianFKStart) to 2019B (\medianFKEnd). More popular websites have less readable policies.

In comparison, Li et al.~\cite{li2012online} measured the reading difficulty of privacy policies for websites for the 30 Dow Jones companies in 2012, finding an average FKGL of 13.33, which is similar to our findings for top 1K and top 10K websites.

\subsection{Privacy policy updates}
We measured the percentage of privacy policies updated in each interval.
To determine if these changes align with shifts in the language of privacy policies, we also measured terms that have experienced changes in their usage at each interval.

\textbf{Privacy policy updates.}
\label{sec:policy-updates}
\begin{figure}[]
\centering
\includegraphics[width=0.99\columnwidth]{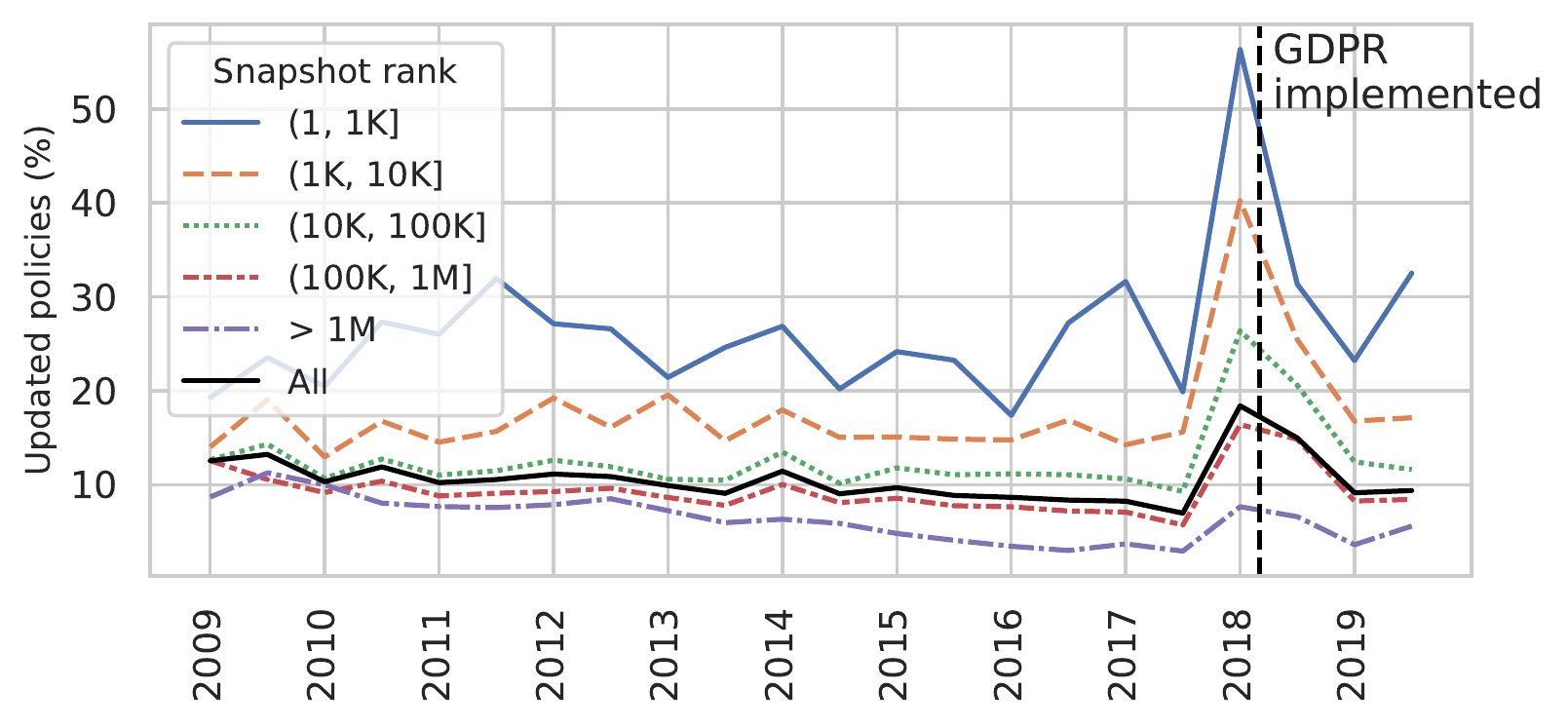}
\caption[Caption]{Percentage of privacy policies updated per interval.\protect\footnotemark}

\Description[Line plot of year vs percentage of policies updated]{While jagged, the trends are relatively stable. Bins closer to 1 have more updated policies. There is a large spike shortly before GDPR is implemented.}

\label{fig:pct-policy-update}
\end{figure}
\footnotetext{We excluded snapshots for which we have a gap in the previous interval from Figure~\ref{fig:pct-policy-update}, since determining the exact time of the updates was not possible.}
Figure~\ref{fig:pct-policy-update} shows the percentage of updated policies, considering
only the~\emph{significant} updates, where the fuzzy similarity ratio~\cite{fuzzywuzzy} of consecutive policy snapshots is less than or equal to 95\%
~\cite{linden2020privacy}.
The spike in 2018 indicates the GDPR's substantial impact on privacy policies.  Although the figure shows that
popular websites update their policies more frequently,
the GDPR appears to have caused a major uptick across all rank buckets. 

\textbf{Change-point concentration.} 
To 
investigate the changes in privacy policy language and vocabulary,
we counted the number of n-gram frequency change-points detected by the PELT algorithm~\cite{killick2012optimal}, using the \textit{ruptures} library~\cite{ruptures}. Change point detection algorithms originate from the signal processing literature, and are designed to identify when a time-series signal has experienced a failure~\cite{picard1985testing}. In our case, n-gram frequency is the signal, and the ``failures'' are events that cause a shift in the usage.

\begin{figure}
    \centering
    \includegraphics[width=0.45\textwidth]{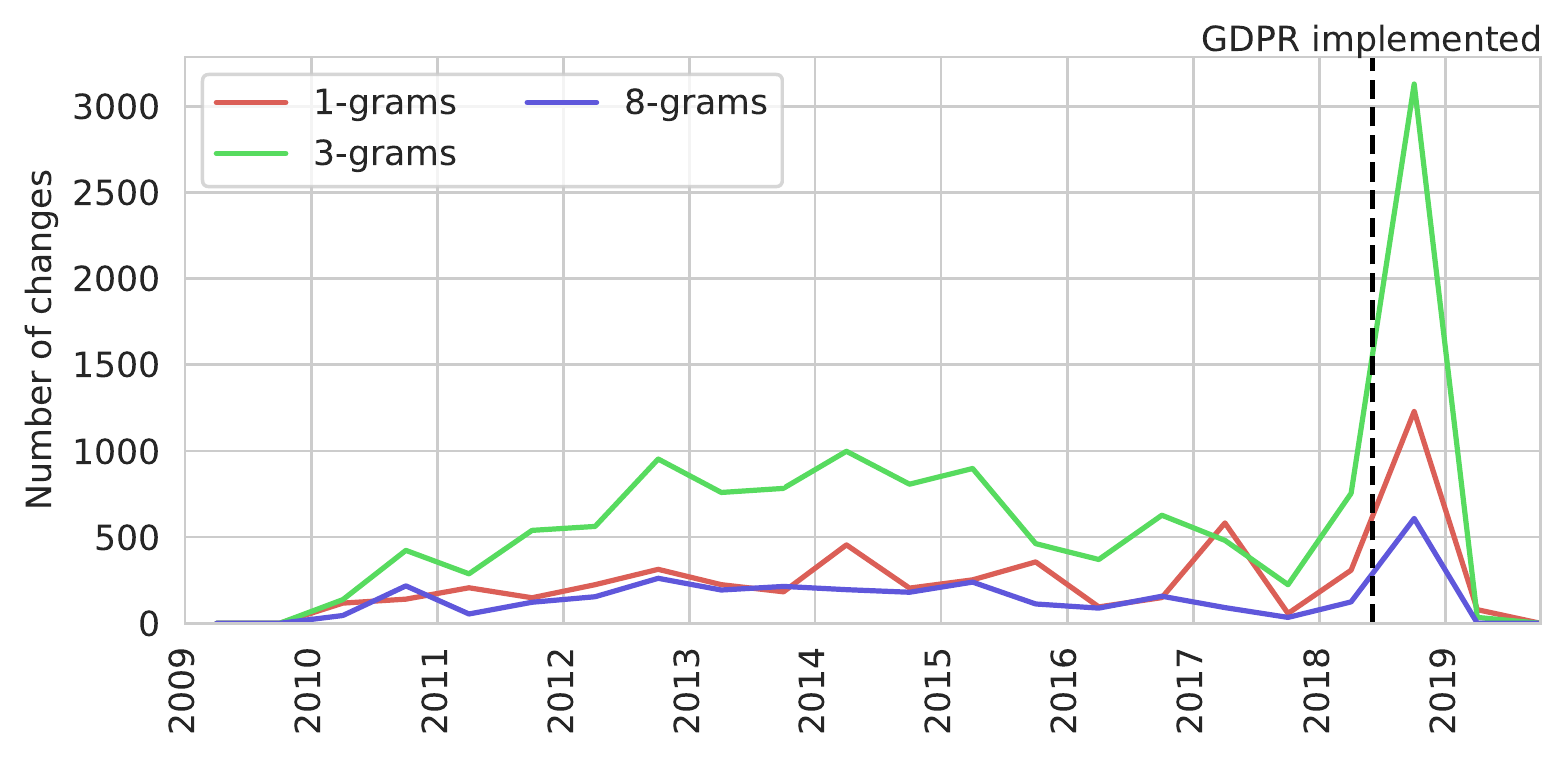}
    \caption{Change point concentration. The y value is the number of frequent n-grams with change points in that interval for 1-grams, 3-grams, and 8-grams.}
    
    \Description[Line plot of year vs number of changes]{There is a roughly arc shaped trend, with 3-grams being the most pronounced. There is a sharp spike right before GDPR is implemented.}
    \label{fig:changepoints}
\end{figure}

We counted the number of change-points at each interval for lemmatized 1-grams and 8-grams with a document frequency of at least $0.01$ in at least one interval.
Figure~\ref{fig:changepoints} shows that the change points for n-grams concentrate around GDPR's introduction, following a similar trend to document level updates in Figure~\ref{fig:pct-policy-update}. This indicates that not only are websites updating their policies, but that the vocabulary used in privacy policies was forced to evolve due to the GDPR.

\textbf{Validation.} To verify that the increase in privacy policy updates we observed in 2018A was related to the GDPR, we took a list of GDPR-related phrases identified by Degeling et al.~\cite{degeling2018we}, plus the phrases “GDPR” and “General Data Protection Regulation,” and selected the subset of 20 phrases with a relative document frequency of less than 1\% in 2015A. We found that documents in 2018A containing at least one of these 20 GDPR-related phrases had a mean update length of 136 new lines, compared to 15 new lines in documents without these phrases, where “update length” is the number of added lines minus the number of deleted lines. The quartile boundaries were at, respectively, 0, 20, 129 and 0, 0, 0. This indicates that policies that contain GDPR-related terms were more likely to have a longer update in comparison to the prior version of the policy, supporting the link between the introduction of the GDPR and the increase in updates in 2018A.

\section{Term-level trends}
\label{sec:analysis}

Motivated by our analysis of updates and change-points, we sought to understand how the language of privacy policies has shifted. While we followed prior research in our use of manually crafted keywords to analyze trends~\cite{degeling2018we},
we additionally developed a system to surface key terms that experts might otherwise overlook. Then we used the key terms surfaced by our tool to guide a more comprehensive analysis into specific concepts.

\subsection{Automated trend surfacing}
\label{sec:automated-trends}

We first built a pipeline to further clean and process the privacy policy texts. We started by replacing certain patterns---URLs, named entities using Spacy~\cite{spacy}, numbers, and emails---in the policies with placeholders. This allowed us to group together phrases that share the same template but include, for example, an organization name. We then extracted terms---n-grams, sentences, named entities, and URLs---for which we obtained the relative document frequency (the percentage of policies they appear in in an interval). Finally, we ranked terms of interest using a variety of scoring functions such as overall gain in relative frequency.

\begin{table}[]
\centering
\resizebox{\columnwidth}{!}{
\begin{tabular}{rrr}
\toprule
\textbf{2-grams (Gain)}        & \textbf{Entities (Pos Slope 2)} & \textbf{2-grams (Pos Slope 2)} \\ \midrule
service providers (0.27) & gdpr (0.12)             & personal data (0.14)    \\
data protection (0.27)   & eu (0.11)               & data protection (0.13)       \\
personal data (0.24)     & google analytics (0.09) & withdraw consent (0.13)      \\
may include (0.24)       & eea (0.08)              & processing personal (0.13)          \\
may collect (0.22)       & facebook (0.07)         & legitimate interests  (0.12)     \\
\end{tabular}
}
\caption{The top five results from three categories of our automated trend surfacing tool. Scoring functions: ``gain,'' is the difference between the lowest and highest frequency; ``pos slope 2,'' is the maximum slope over any two intervals.}
\label{tbl:automated_results}
\end{table}

\textbf{Results.} Our trend surfacing tool returned hundreds of phrases which we manually inspected. 
We provide some examples of top-scoring trends that our tool identified in Table \ref{tbl:automated_results}. Under 2-grams measured by gain, we note the growth of terms such as ``may include'' and ``may collect,'' which are terms of ambiguity in privacy policies (previously studied by Reidenberg et al.~\cite{reidenberg2016ambiguity}). This result suggests that privacy policies may be becoming more ambiguous over time, which should be investigated in future work. We noticed growth in terms related to European privacy regulation (GDPR, EU, EEA), along with Google Analytics and Facebook, two commonly used third-party services. We investigate this trend and other trends we observed in more depth in Section~\ref{subsec:trends}. Furthermore, we note growth in terms that likely relate to GDPR.

\subsection{Trends}
\label{subsec:trends}

After examining the results of our automated trend surfacing tool, we chose to investigate trends in three categories: self-regulatory bodies, tracking technologies, and third parties. We investigated tracking technologies because our tool identified several dozen terms of interest related to cookies and web beacons. Similarly, we included third parties as a category because our tool identified terms including Google, Facebook, Instagram, and Twitter. Furthermore, adequate disclosure in each of these categories is crucial to demonstrate the transparency that is a premise of \textit{notice and choice}. Finally, we chose to investigate self-regulatory initiatives because our tool identified several terms containing TrustArc, NAI, and DAA as gaining traction, which may be of interest to regulators.

\begin{figure}
    \centering
    \includegraphics[width=0.45\textwidth]{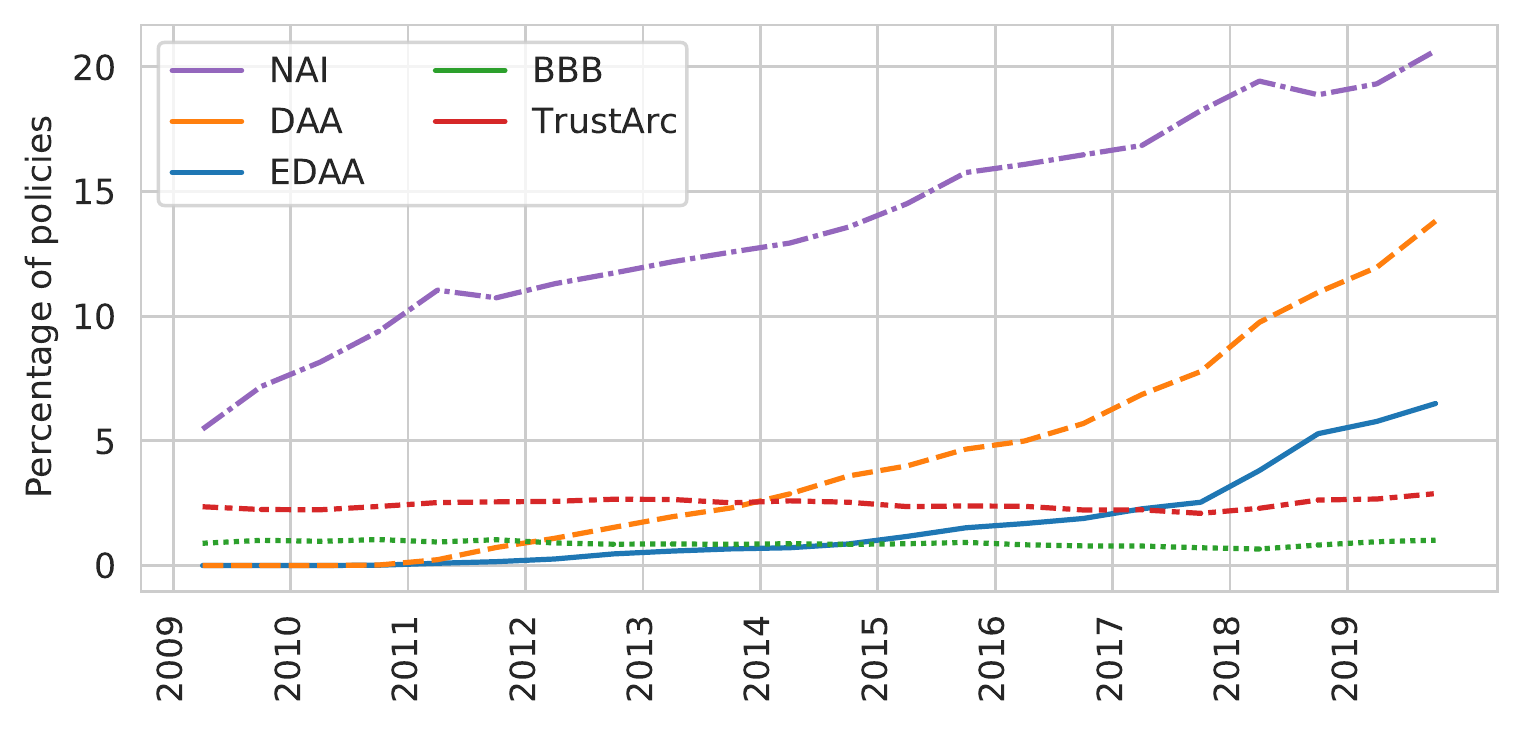}
    \caption{The percentage of policies, after deduplication, referencing self-regulation bodies over time. }
    
    \Description[Line plot of year vs percentage of policies containing the trend]{The following terms trend roughly linearly upwards, in descending order of final value: NAI, DAA, EDAA. The following terms trend roughly flat: BBB, TrustArc.}
    \label{fig:selfreg}
\end{figure}

\textbf{Self-regulatory initiatives.} 
We began by identifying a set of self-regulatory initiatives. We chose six privacy seal vendors from prior work~\cite{rodrigues2013developing} and three advertising industry trade groups. We then measured how often privacy policies referred to these self-regulatory initiatives, counting a privacy policy as referencing an initiative if it either included the initiative's name or a link to the initiative's website.\footnote{We combined TrustArc and TRUSTe in our analysis, since they are the same entity. Our automated trend detection identified TrustArc as a term with increasing frequency, but when combined with TRUSTe the frequency of references is stable over time.} Figure~\ref{fig:selfreg} plots these references over time, omitting the initiatives that were mentioned in fewer than 1\% of privacy policies.

The graph indicates that self-regulatory online advertising trade groups---the Network Advertising Initiative (NAI), Digital Advertising Alliance (DAA), and European Interactive Digital Advertising Alliance (EDAA)---have seen substantial growth in mentions in privacy policies. The EDAA appears to have more than doubled its presence after the GDPR's introduction, and the NAI has risen around four-fold in 10 years.
By contrast, self-regulatory initiatives that are more commonly associated with first-party websites---
TrustArc and the Better Business Bureau (BBB)---have remained relatively stagnant over time.

\begin{figure}[t]
    \centering
    \includegraphics[width=0.45\textwidth]{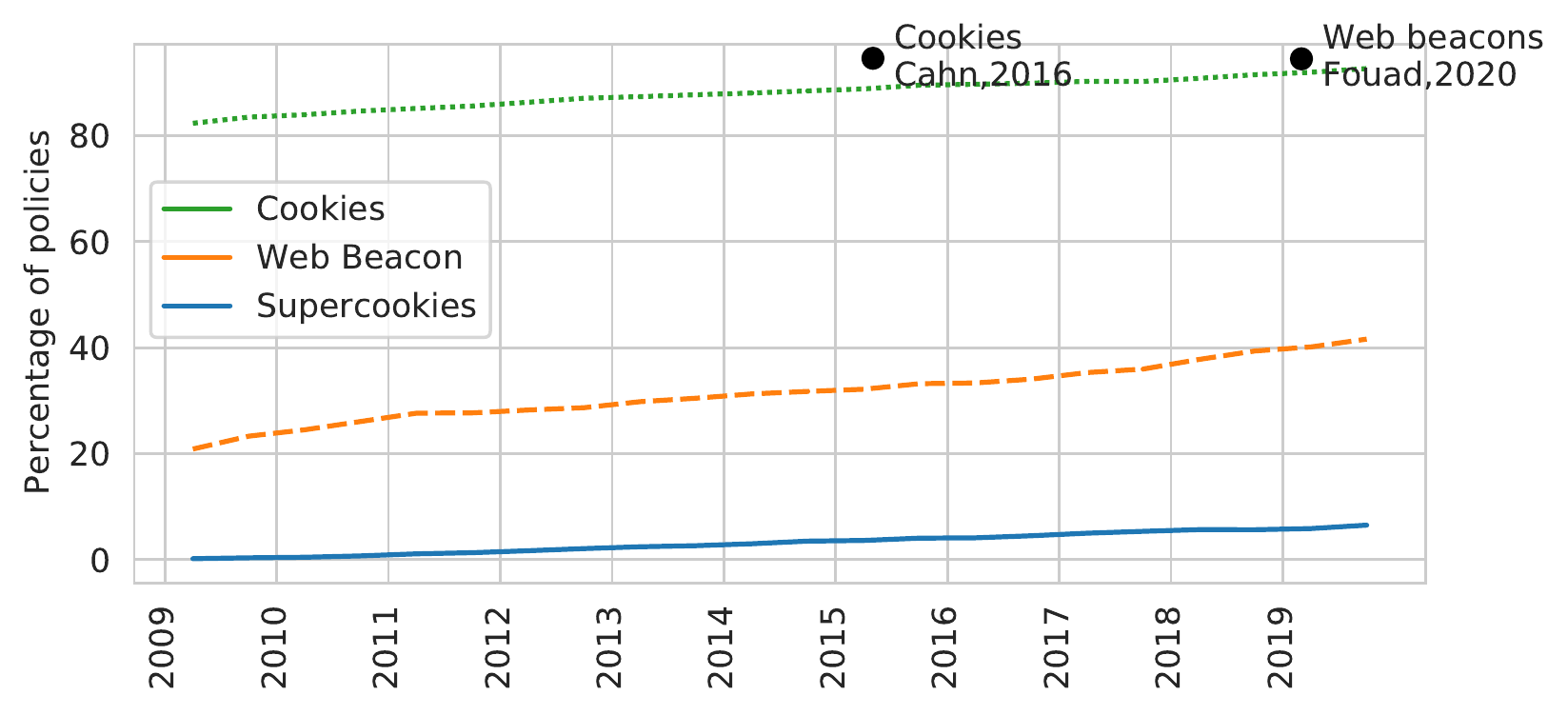}
    \caption{The percentage of policies, after deduplication, that reference terms related to stateful tracking technologies. The dot labeled ``Cookies'' is the percentage of websites in the Alexa top 100K with any observed cookies~\cite{cahn2016empirical}. The dot ``Web beacons'' is the percentage of websites in the Alexa top 10K with observed pixels~\cite{Fouad2020missed}.}
    \Description[Line plot of year vs percentage of policies containing the trend]{The following terms trend slightly upwards, in descending order of final value: cookies, Web Beacon, Supercookies. Other researchers measurements for cookies approximately match the value at that point.}
    \label{fig:trackingtech}
\end{figure}

\begin{figure}[t]
    \centering
    \includegraphics[width=0.45\textwidth]{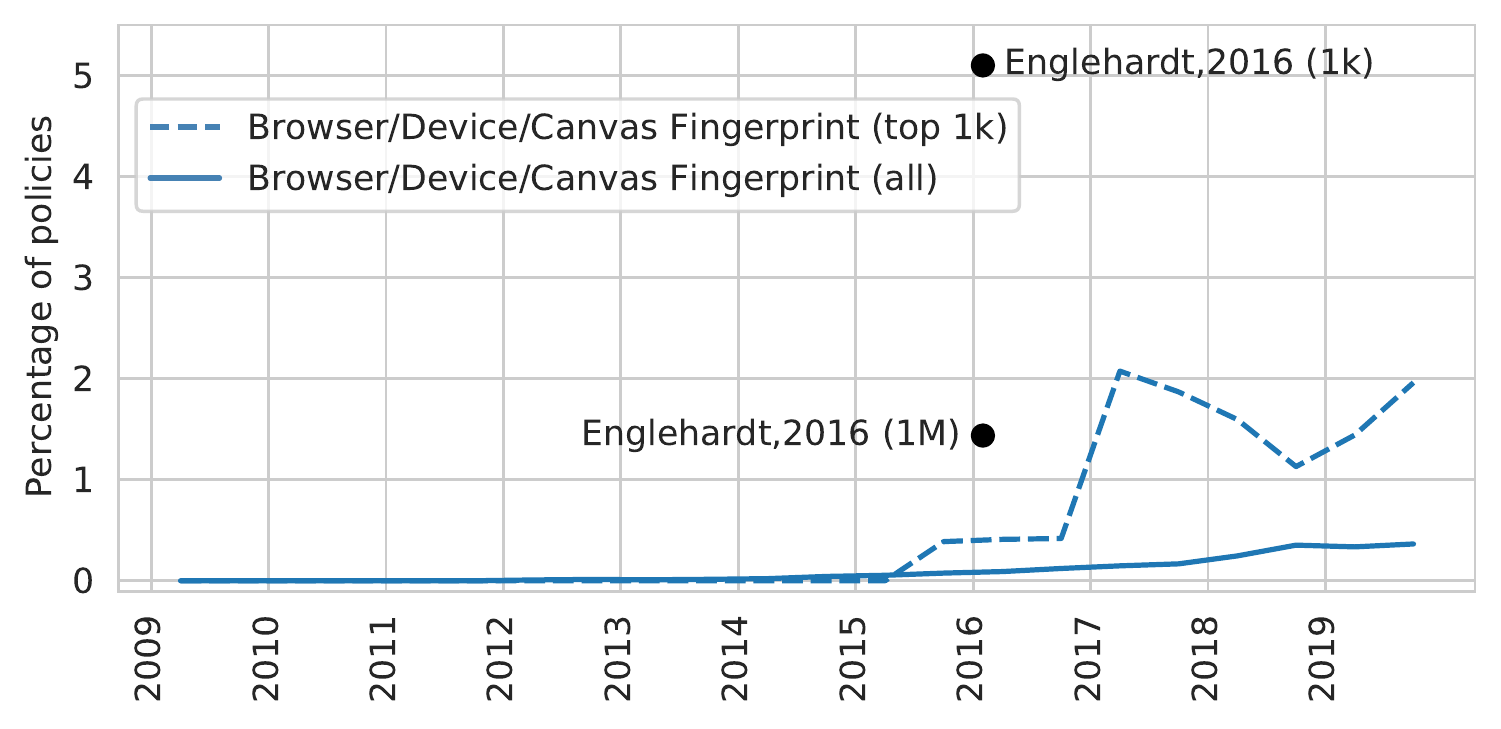}
    \caption{The percentage of policies, after deduplication, that reference terms related to fingerprinting. The dot labeled ``1K'' is the percentage of websites in the Alexa top 1K at the time with observed canvas fingerprinting, and ``1M'' is the top 1M~\cite{englehardt2016online}.}
    \Description[Line plot of year vs percentage of policies containing the trend]{This graph shows Browser/Device/Canvas fingerprinting for all against the top 1k. The top 1k grows faster than all, peaking, then dropping, one year after Englehardt et al., 2016.}
    \label{fig:fingerprinting}
\end{figure}

\textbf{Tracking technologies.} Tracking technologies are pervasive in the modern web~\cite{englehardt2016online}. If privacy policies were fully transparent, we would expect that observations of tracking technologies on the web would match mentions in privacy policies, so we investigated how frequently privacy policies disclose the use of tracking technologies. 

Using our automated trend surfacing tool as a guide, we selected four terms related to tracking technologies, which we split into two categories: stateful tracking technologies and stateless (``fingerprinting'') tracking technologies~\cite{mayer2012}. We constructed regular expressions to match all names listed in the respective Wikipedia entry for the technology~\cite{wikipediaCookie,wikipediaWebBeacon,wikipediaDeviceFingerprint,wikipediaCanvasFingerprint,wikipediaLSO,wikipediaEvercookie}.
The percentage of privacy policies containing the first and second
groups are shown in Figure \ref{fig:trackingtech} and Figure~\ref{fig:fingerprinting}, respectively. Mentions of ``cookie'' are quite close to the observations of the usage of cookies by Englehardt and Narayanan~\cite{englehardt2016online}, and Cahn et al.~\cite{cahn2016empirical}. However, web beacons and fingerprinting appear to be underreported. Fouad et al. measured that 94.6\% of the Alexa top 10K websites had web beacons~\cite{Fouad2020missed}, compared to 25.8\% of policies in 2019 mentioning the related terms. Further, fingerprinting related terms were mentioned far less commonly than Englehardt and Narayanan's measurements of canvas fingerprinting in both the top 1K and top 1M~\cite{englehardt2016online}.

\begin{figure}
    \centering
    \includegraphics[width=0.45\textwidth]{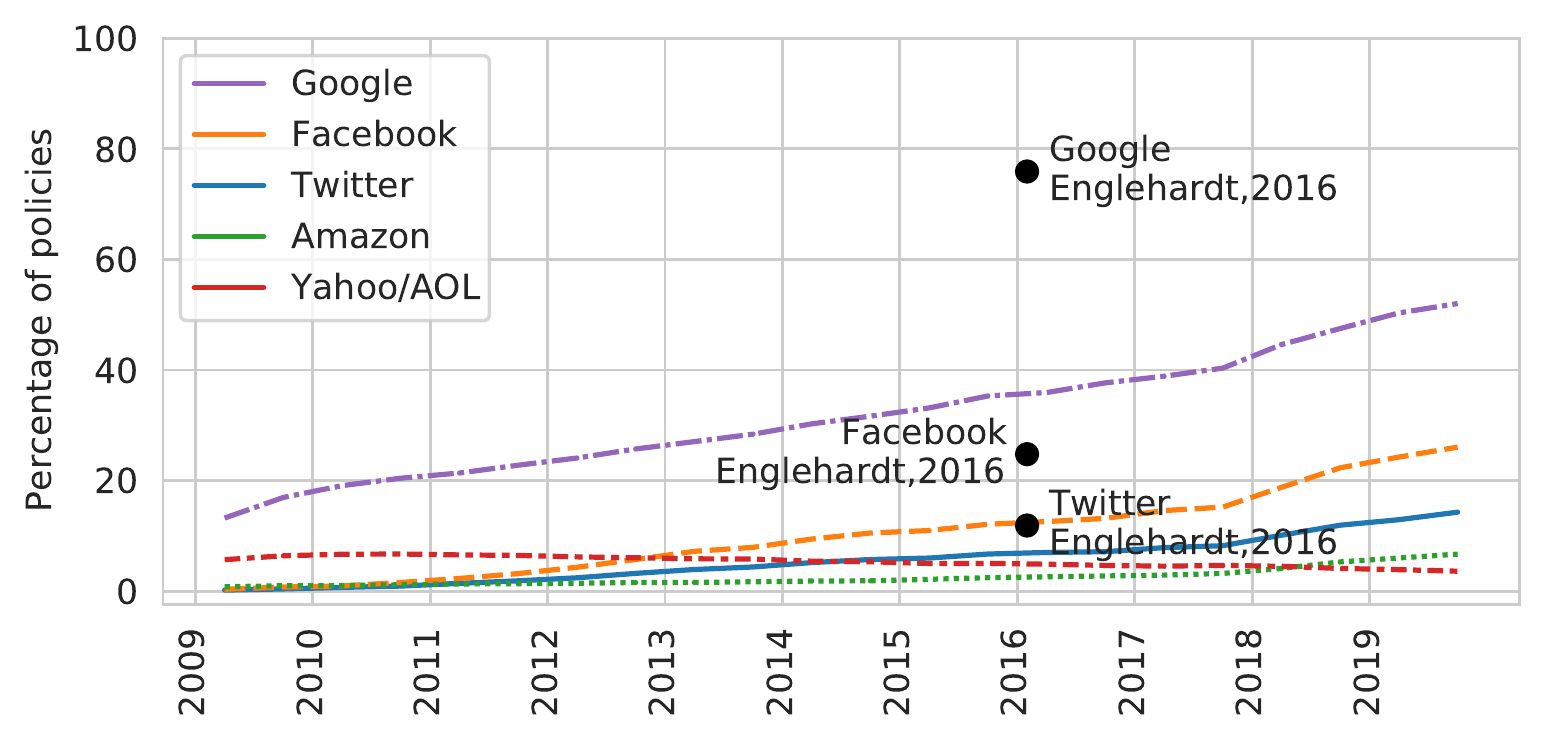}
    \caption{The percentage of policies, after deduplication, that reference specific third parties. A match for a common name for the third-party organization, or a match in a link for any domain owned by the third-party organization or their parent organization was considered a reference. Dots indicate measurements of the presence of third parties as a tracker~\cite{englehardt2016online}.}
    \Description[Line plot of year vs percentage of policies containing the trend]{The following terms trend upwards, in descending order of final value: Google, Facebook, Twitter. The following trends remain flat: Amazon, Yahoo/AOL. Measurements for Twitter from Englehardt et al. are larger but somewhat close to policy mentions. Measurements for Facebook and Google are substantially larger.}
    \label{fig:trackers}
\end{figure}

\textbf{Third parties.} Third-party services are often associated with tracking users on the web~\cite{englehardt2016online}. Continuing our investigation of disclosure and transparency, we chose to examine how often privacy policies disclose the prevalence of third parties.

We compiled a list of 20 popular third parties from other measurements of third parties on the web~\cite{trackerradar, englehardt2016online,lerner2016internet}. We searched for alternative names for each third party and domain names operated by the third party using data from DuckDuckGo's Tracker Radar project~\cite{trackerradar}. We chose the top five most referenced third parties in 2019B, which we show in Figure~\ref{fig:trackers}. We found that these third parties were mentioned in privacy policies far less frequently than they were observed in the web measurements~\cite{trackerradar,englehardt2016online}. 

The gap between observation and disclosure indicates that many privacy policies may not be disclosing all present third parties. 
We note that reporting of third parties is largely increasing. It is unclear if this is due to increased presence or transparency.

\textbf{Validation.} We evaluated our matching criteria for tracking technologies, third parties, and self-regulatory initiatives, to ensure that we were studying the intended trends. Table \ref{tab:selfreg-NDE-labeled} presents results from manually labeling 100 positive examples of each matching category, showing that our matching criteria have high precision.

{
\begin{table}[]
\centering
\resizebox{0.8\columnwidth}{!}{%
\begin{tabular}{@{}lcc@{}}
\toprule
\textbf{Query} & \textbf{Positive} & \textbf{Negative} \\ \midrule
Cookies & 99 & 1\\
Web Beacon & 89 & 11\\
Supercookies & 99 & 1\\
Browser/device/canvas fingerprint& 100 & 0\\
\midrule
Google & 95 & 5\\
Facebook & 86 & 14\\
Twitter & 73 & 27\\
Amazon & 86 & 14\\
Yahoo/AOL & 97 & 3\\
\midrule
TrustArc & 87 & 13\\
BBB & 97 & 3\\
NAI & 95 & 5\\
DAA & 100 & 0 \\
EDAA & 100 & 0\\
\bottomrule
\end{tabular}%
}
\caption{Manual validation of 100 positives for each query. For third parties and tracking technologies, positives indicate the term is used in a context related to tracking. For self-regulatory initiatives, positives indicate a relationship with the initiative.}
\label{tab:selfreg-NDE-labeled}
\end{table}
}

\begin{table}[]
\centering

\resizebox{\columnwidth}{!}{
\begin{tabular}{@{}lrl@{}}
\toprule
\textbf{Privacy policy URL} & \textbf{\# of websites with linking policy} &  \\ \midrule
google.com/privacy\_ads.html & 11,324 &  \\
google.com/intl/en/policies/privacy & 1,690 &  \\
google.com/policies/privacy & 1,421 &  \\
automattic.com/privacy & 948 &  \\
twitter.com/privacy & 931 &  \\
google.com/privacy.html & 873 &  \\
facebook.com/policy.php & 607 &  \\
google.com/privacypolicy.html & 559 &  \\
mailchimp.com/legal/privacy & 528 &  \\
doubleclick.net/us/corporate/privacy & 498 &  \\ \bottomrule
\end{tabular}
}
\caption{Privacy policies that are most linked from the other privacy policies. The right column indicates the distinct number of sites.
}

\label{tab:most-linked-policies}
\end{table}
\textbf{Outbound links.} In addition to disclosing third-party partners, companies may provide links to opt-out pages, industry organizations, or third-party partners that they share data with. Analyzing links found in the privacy policies may reveal trends in third-party data sharing and the effect of regulation on transparency. Here we study links from one privacy policy to another.

We found that 20.3\% of websites link to one or more additional privacy policies. 
This result implies that users wishing to achieve a comprehensive understanding of a website's data practices face an increased burden of reading both the first-party policy and the linked policies.
We show the ten privacy policy URLs with the most incoming links from other policies in Table \ref{tab:most-linked-policies}, which includes six policy URLs from Google. 
  
The privacy links we observe provide an alternate explanation to the disparity we observed in our discussion of tracking technologies
---specific technologies may not be named because they are explained in the third party's privacy policy---instead of the first party's.

\section{Conclusion}
\label{sec:conclusion}
We developed a system for the longitudinal, large-scale, automated collection and curation of privacy policies. Using the system we built a dataset of over 1M privacy policies that span more than two decades. We found that privacy policies are becoming longer and harder to read.

We developed an automated trend surfacing tool and investigated some of the surfaced trends. Specifically, we investigated trends around third parties, tracking technologies, and self-regulatory bodies. Our results suggest that privacy policies show a concerning lack of transparency: the usage of third parties and tracking technologies is severely underreported.

Our results add to the growing body of evidence suggesting the inadequacy of the ``notice and choice'' model for privacy policy regulation and demonstrate the monumental impact of the GDPR.

\textbf{Release.} Our dataset and source code is available for public use.\footnote{ \url{https://privacypolicies.cs.princeton.edu/}} The access requests we have received have revealed a diversity of use cases for our data, including the study of responses to regulation, automated enforcement, health information technology, and ethics.

In addition to making data and code available, inspired by the TOSBack project~\cite{TOSBack}, we built a chronologically accurate GitHub repository of the policy documents in our dataset.
We used \emph{GitPython}~\cite{GitPython} to commit privacy policy texts and HTML sources (in a separate branch) using the time-of-archive timestamps and detailed privacy policy metadata. 
GitHub’s easy to use web interface enables users without technical skills to perform full text search, compare different versions of the same policy, or use the ``Blame’’ feature to trace modifications.
Finally, we built a separate web interface
to make it easier to search, filter, and explore the privacy policies in the GitHub repository.

\textbf{Limitations.}
Our data collection and analyses were limited to privacy policies; we did not analyze how archived websites track users or share personal data with third parties. 

For internationalized websites, the location of the Wayback Machine crawlers could determine the language of the archived site and its privacy policy. 
Although ``the vast majority of captures are initiated from the [United States],'' there is no way to determine the location, country, or IP address of the Wayback Machine crawlers that captured a particular snapshot~\cite{crawl-location-tweet-by-graham}.

For our readability discussion, we noted that readability metrics have limitations; they do not consider organization, formatting, or length of the document, nor the semantic difficulty of the document. 

In our discussion of how terms trend over time, we used (typically handcrafted) regular expressions to search for referenced terms. Despite our best efforts and extensive manual validation, these regular expressions may capture terms that we did not intend to capture. They additionally may capture terms we intended to capture, but do not fit the targeted concept, and they may miss terms that fit the targeted concept.

\textbf{Future work.}
Based on our results we suggest several directions for future work. The reporting gap for third parties and tracking technologies could be investigated with a combination of web measurements and privacy policy analysis. Another direction is to further investigate ambiguity in privacy policies.

More sophisticated natural language processing techniques applied to longitudinal privacy policy data could improve our understanding of how online services adapt to new privacy regulations, the prevalence of specific business practices, and whether online services are complying with privacy laws. 
We anticipate that our dataset will prove useful for research in all of these directions.

\vspace{-0.06cm}
\begin{acks}
We thank Mohamed El-Dirany for his help with the initial exploration of this space, Prateek Mittal for his advice and guidance, Lucina Schwartz for help with editing, and the reviewers for their insightful comments.
Gunes Acar holds a Postdoctoral fellowship of the Research Foundation Flanders (FWO). This work was supported by CyberSecurity Research Flanders (VR20192203).
\end{acks}
\bibliographystyle{ACM-Reference-Format}
\bibliography{ref.bib}


\end{document}